\documentstyle[psfig,amssymb,graphicx,epsf,latexsym,longtable,supertabular,rotating]{mn}     

\title[Spectroscopy of SN 1987A at 0.9--2.4~$\mu$m:  Days 1348-3158]
{Spectroscopy of SN 1987A at 0.9--2.4~$\mu$m: Days 1348-3158}
\author[A.Fassia {\it et al.}]
{A. Fassia$^{1}$, W.P.S. Meikle$^{1}$ and J. Spyromilio$^{2}$
\\
\\
$^1$Astrophysics Group, Blackett Laboratory, Imperial College, Prince
Consort Rd, London SW7 2BZ, UK\\
$^2$European Southern Observatory, Karl-Schwarzschild-Strasse 2,
Garching,  Germany\\  
}
\begin{document}
\maketitle  
\begin{abstract}
We present near-infrared spectroscopic observations of SN~1987A
covering the period 1358 to 3158 days post-explosion.  This is the
first time that IR spectra of a supernova have been obtained to such
late epochs.  The spectra comprise emission from both the ejecta and
the bright, ring-shaped circumstellar medium (CSM). The most prominent
CSM emission lines are recombination lines of H~I and He~I, and
forbidden lines of [S~III] and [Fe~II].  The ejecta spectra include
allowed lines of H~I, He~I and Na~I and forbidden lines of [Si~I],
[Fe~I], [Fe~II], and possibly [S~I].  The intensity ratios and widths
of the H~I ejecta lines are consistent with a low-temperature Case~B
recombination spectrum arising from non-thermal ionisation/excitation
in an extended, adiabatically cooled H-envelope, as predicted by
several authors.  The slow decline of the ejecta forbidden lines,
especially those of [Si~I], indicates that pure non-thermal excitation
was taking place, driven increasingly by the decay of $^{44}$Ti.  The
ejecta iron exhibits particularly high velocities (4000--4500~km/s),
supporting scenarios where fast radioactive nickel is created and
ejected just after the core-bounce.  In addition, the ejecta lines
continue to exhibit blueshifts with values $\sim$--200~km/s to
--800~km/s to at least day~2000. These blueshifts, which first
appeared around day~600, probably indicate that very dense
concentrations of dust persist in the ejecta, although an alternative
explanation of asymmetry in the excitation conditions is not ruled
out.\\
\end{abstract}

\section{Introduction}
The exceptionally close proximity of the type~IIpec supernova SN~1987A
in the Large Magellanic Cloud has provided a unique opportunity to
observe a core-collapse supernova with all the resources of modern
observational astronomy and over a very long period of time.
Near-infrared (near-IR) spectroscopy has played a vital role in the
determination of the physical conditions in the debris and the
investigation of the element synthesis, providing a valuable
complement to optical spectra.  In particular, in the nebular phase
(when the lines are mostly optically thin) the line profiles produced
in the high velocity, homologous expansion have enabled us to examine
the ejecta abundances and their {\it spatial} distribution.  During
the first few years, all the major southern observatories obtained
near-IR spectra of SN~1987A.  Observations at the Anglo-Australian
Telescope (AAT) covering the first 3~years post-explosion were
described in Meikle et al. (1989, 1993) (hereinafter Papers~I, II). \\

\noindent In this paper we describe spectra obtained at the AAT for a
further 8 epochs from 1990 November 2 (day 1348 = 3.7~years) to 1995
October 17 (day 3158 = 8.6~years). Epochs are with respect to the
explosion date 1987 February 23.  Our day~1348 spectrum was the last
observation of SN~1987A obtained with the near-IR spectrometer FIGS
(Bailey et al. 1988.  Also see Papers~I \& II). All the
subsequent spectra were obtained with the more sensitive spectrograph,
IRIS (Allen et al. 1993), which allowed the acquisition of high
resolution near-IR spectra to a later phase than was achieved at any
other observatory.  Preliminary reports of these data have been given
in Fassia (1999) and Meikle (2001).  The only other post-3~year IR
spectra reported are those of Bautista et al. (1995) which reached
day~1445, but at a lower resolution than those described here.  In a
future paper (Fassia et al., in preparation) we shall present $HK$-band
spectra of SN~1987A taken with the MPE imaging spectrograph, 3D
(Weitzel et al. 1996), at the AAT on 1997 Dec 16 (day~3949 =
10.8~years) and 1998 Dec 1 (day~4299 = 11.8~years).  We also note that
the earliest Hubble Space Telescope (HST) NICMOS spectrum of SN~1987A
was taken on 1998 June 15 (day~4130 = 11.3~years), but has not
yet been published.

\section{Observations}
The observations are summarised in Table~1.  Some of the data
presented here were obtained in runs of several days. In the text and
diagrams these runs will usually be identified by the epoch of the
first day of each.  For further details of the FIGS spectrometer, used
to obtain the day~1348 spectrum, see Bailey et al. (1988) and
Papers~I \& II.  The IRIS spectrograph was based on a $128\times128$
pixel HgCdTe array.  All the IRIS observations were taken at f/36,
yielding a plate scale of 0.79 arcsec/pixel.  Spectra were acquired in
the echelle mode for epochs 1469--2112~days and in the $H$-grism mode
for the subsequent epochs.  In the echelle mode, 4 orders of
cross-dispersed spectra were provided, covering 0.9--1.5~$\mu$m
($IJ$~band) and 1.4--2.4 $\mu$m ($HK$~band).  The slit was usually
oriented at a P.A. of 90 deg. {\it i.e.} east-west.  The slit length
was 13 arcsec (16 pixels) while in the dispersion direction, 2 pixels
corresponded to a spectral resolution of about 375
($\lambda/\Delta\lambda$) (equivalent to $\sim$800~km/s). However, the
actual resolution was generally poorer than this (see below). To
enable sky-subtraction, the telescope was nodded along the slit by
5--8~arcsec (6--10 pixels).  To improve the flux measurement precision
of both the ejecta and the bright ring, the slit width for the $IJ$
observations was set at 5.8~arcsec. Similarly, the 1348~d FIGS
spectrum was obtained with a 5.9~arcsec aperture. Thus, the
resolution of the (point-source) ejecta spectra was effectively
determined by the seeing.  For the CSM spectra, the finite extent of
the bright ring ($1.66\times1.21$~arcsec, Plait et al. 1995)
would have produced a slightly lower resolution.  The $IJ$ echelle
resolutions listed in Table~1, column~4, were obtained directly from
the FWHM of the He~I 1.083~$\mu$m line.  They were typically 200--400
($\lambda/\Delta\lambda$). As already mentioned, the resolution of the
ejecta spectra was probably slightly higher.  For the $HK$
observations, we were unable to use the 5.8~arcsec slit since it
produced severe degradation of the S/N. We therefore used a slit width
of 1.4~arcsec in this band. The $HK$ resolution was obtained from the
FWHM of the He~I 2.058~$\mu$m CSM line.  Owing to the declining flux,
after day~2112, spectra were taken using the lower resolution
$H$-grism mode. This covered 1.2--2.1~$\mu$m in a single order.  The
slit was usually oriented at a P.A. of 135 deg. in order to maximise
the spatial separation of the spectra of the supernova and Stars 2 and
3 (see below). The slit length was 24~arcsec (30~pixels), while in the
dispersion direction a 1.6~arcsec (2 pixel) wide slit was used,
yielding a spectral resolution of 120 ($\lambda/\Delta\lambda$)
(equivalent to $\sim$2,500~km/s).  The nod throw was 10--15~arcsec. \\

\noindent The day~1348 FIGS spectrum was reduced using FIGARO
(Shortridge 1995).  Final fluxing was by comparison with contemporary
broad-band photometry obtained with the AAT IR photometer IRPS ({\it
cf.}  procedure described in Paper~II), using the A7 dwarf BS~2015 as
the local standard.  However, by this epoch, the narrow He~I
1.083~$\mu$m line from the bright ring had an intensity of about 20\%
of the flux within the $J$-photometry wavelength window.  Moreover it
lay right at the blue edge of the $J$-band filter, making it difficult
to judge the effect of the bright ring on the $J$-magnitude. (The
spatial coverage of IRPS included the ring).  Therefore, taking into
account other uncertainties, we judge the uncertainty in the fluxing
to be of order $\pm$25\%.  The wavelength scale was calibrated using
the narrow He~I 1.083~$\mu$m and Pa$\beta$ lines, adopting a redshift
of +289~km/s (Crotts \& Heathcote 1991). \\

\noindent The IRIS data were also reduced using FIGARO.  After
bias-subtraction and flat-field correction the image-pair obtained at
the two nod positions were subtracted to remove sky line emission. For
the echelle-mode data, different echelle orders were then traced and
corrected for echelle distortion. The positive and negative spectra
were extracted from the resulting frames using simple extraction.  We
extracted over 4--6 pixel rows (3.2--4.8~arcsec), depending on the
seeing.  This is discussed further below.  Cosmic rays and residual
sky-lines were identified and removed by comparing repeat observations
of spectra obtained at different times during the night.  Wavelength
calibration was by means of argon and xenon arc lamps and the
night-sky emission lines present in the supernova spectra. The
wavelength uncertainty in the echelle spectra ranges from
$\pm$0.0014~$\mu$m in the $IJ$-band spectra to $\pm$0.0025~$\mu$m in the
$HK$-band spectra. The wavelength uncertainty in the grism spectra is
$\pm$0.009~$\mu$m.  To correct for the atmospheric and instrumental
transmission functions and to flux-calibrate the spectra, we used the
G~dwarf spectrophotometric standard BS~1294 (Allen \& Cragg 1983).
The adopted magnitudes were J=+5.20, H=+4.90, K=+4.84.  (For the grism
spectrum on day~3158, we used the A4~giant HD 19904, adopting
H=+6.66). The spectral orders of the echelle spectra were then merged
to form single spectra. The final fluxing of the spectra is now
described. \\

\noindent Accurate flux calibration of the spectra was difficult and
became increasingly so with time.  This was due to the combined
effects of the declining SN flux, variable seeing, variable
atmospheric transmission for both supernova and standard (the airmass
was inevitably quite high {\it viz.} 1.4--1.8), pointing errors of up
to $\sim$2~arcsec and, for the echelle spectra, the small number of
detector pixels (16) along the slit in a given order.  Flux
measurement under poor seeing conditions was a particular problem for
the narrow slit spectra due to the greater (but uncertain) vignetting
effects.  However, the most difficult problem of all in the flux
calibration procedure was contamination by light from the nearby
Star~3, and even from the slightly more distant Star~2.  To make
matters worse, Star~3 is a variable Be star.  Around day~1500 the IR
fluxes from Stars 2 and 3 were about $\times$1.4 and $\times$0.85,
respectively, of the SN flux, rising to about $\times$4 and
$\times$1.5 by day~2112 (Walborn et al. 1993, Suntzeff private
communication. See also Turner et al. (1996)).  Star~2 lies at
2.9~arcsec from the SN, at PA = $311^{\circ}$ while Star~3 lies at
1.6~arcsec distance at PA = $117^{\circ}$ (Walker \& Suntzeff 1990).
Thus, for the typical E-W slit orientation used for echelle spectra,
the distances of Stars~2 and 3 from SN~1987A, along the slit, were
only $\sim$2.2 and $\sim$1.5~arcsec respectively.  (We note that even
at our latest epoch, the intrinsic diameter of the ejecta source was
probably less than 0.1~arcsec (Jakobsen et al. 1994), and so was
effectively a point source whose observed extent was entirely
determined by the seeing.)  Thus, under typical seeing conditions, the
Star~3 spectrum was spatially blended with the spectra of both the
ejecta and bright ring of SN~1987A, while under poor seeing conditions
even Star~2 could significantly contaminate the SN light.  The grism
spectra of days~2952 and 3158 were taken with a slit PA of
135$^{\circ}$ yielding slightly larger displacements of the stars
along the slit.  However, the much reduced flux from the supernova by
these epochs meant that contamination of the SN light by the nearby
stars was still a problem.  The contamination from both stars took the
form of a continuum of uncertain flux, together with possible narrow H
and He emission lines from Star~3 (Walborn et al.  1993). \\

\noindent Assessment and removal of the contaminating light presented
a difficult challenge. Even with the wide slit, variations in the
seeing conditions and/or telescope pointing precision could make the
degree of contamination within the slit difficult to estimate.  (A
further complicating factor was that Star~3 itself could vary by up to
a magnitude over a period of 800 days.  However, the IR light curves
of the star provided by (Walborn et al. 1993) meant that this
variation could, in principle, be taken into account.)  As already
described, we extracted the spectra from 4--6 pixel rows
(3.2--4.8~arcsec).  Thus all the extracted spectra contained
contamination from Star~3 and sometimes Star~2. \\

\noindent We first consider the removal of the continuum contamination
from the ejecta spectra.  We investigated several methods using
ground-based observations.  However, internal consistency in the
results eluded us, probably due to uncertainties in the variable
factors already described.  Therefore, we decided that the best
solution was to make use of the high spatial resolution optical
spectra provided by the HST. To determine the extent of the continuum
contamination, we inspected the relatively uncontaminated optical HST
spectra of the SN~1987A ejecta obtained on days 1862, 2210 (Wang et
al. 1996) and 2875 (Chugai et al. 1997 (C97)). The red end
(6000--8000~\AA) of these spectra exhibits a weak, roughly flat
continuum of $1-2\times10^{-17}$ erg sec$^{-1}$\AA$^{-1}$.  C97 and
DeKool, Li \& McCray (1998) (dKLM98) suggest that this is actually a
quasi-continuum produced by the down-conversion of UV photons
following absorption/emission by metals.  We therefore assumed that
the ejecta continuum in the near-IR bands should also be roughly flat.
However, in the reduced IR spectra (before correction for stellar
contamination), the continuum was observed to rise towards shorter
wavelengths and so we attributed the rising portion to contamination
from Stars~2 \& 3.  To remove the contamination, we first modelled the
shape of the stellar continuum by fitting a blackbody of the
appropriate temperature to the Star~3 infrared photometry of Walborn
et al.  (1993). Since Stars~2 \& 3 are of a similar stellar type (B2)
the model continuum was used to represent the combined effects of
contamination from both stars.  However, Star~3 was the dominant or
even sole contaminator in all cases.  We then scaled the model
continuum and subtracted it from the reduced spectrum to try to
produce a flat continuum.  For most of the wavelength range covered,
it was found that roughly flat continua could indeed be formed in this
way.  However, within the $\sim$1.4~$\mu$m--$\sim$1.7~$\mu$m range a
relatively abrupt decline, or ``step'' in the continuum persisted.  We
believe this step to be due to a real physical decline in the
quasi-continuum of the ejecta, and is not due to the fact that the
$IJ$, $HK$ spectra were taken at different times and with different
instrument settings.  Our reasons are, firstly, the $IJ$ spectra
covered wavelengths up to 1.5~$\mu$m and the beginning of the ``step''
is clearly present in these data alone.  Secondly, the step can be
seen in the spectra taken at the AAT (Paper~II) and at CTIO (Elias et
al. 1991) at $\sim$1000~days, and in the day~1445 spectrum of Bautista
et al.  (1995). \\

\noindent To finally flux-calibrate the de-contaminated spectra we
proceeded as follows. First we edited out the narrow lines, producing
observed ``pure-ejecta'' spectra, $S_{obs.}(\lambda)$.  For the
low-resolution grism spectra, this could only be done very
approximately.  However, the CSM line contribution to the flux in the
wavelength region covered by the grism (1.2--2.1~$\mu$m) was probably
small.  This can be deduced by considering the CSM line fluxes
(Table~3) at earlier epochs (days~1469--2112) when these lines were
more clearly resolved.  Comparison with contemporary pure-ejecta
$JHK$~band magnitudes obtained at CTIO (Suntzeff private
communication) or with the empirical spectral model matches
(Figs.~3--5) (see below) indicates that the CSM contribution was
probably less than 20\% of the total during this earlier phase, and
showed little sign of increasing with time.  We believe it is highly
unlikely that a sudden rapid increase in the CSM contribution occurred
by day~$\sim$3000.  \\

\noindent We then scaled the individual $JHK$ regions of the spectra
to match the contemporary CTIO photometry.  To find the scaling
factors, we determined the true ejecta spectrum, $S(\lambda)$, using
the relation

\begin{equation}
mag=-2.5\log\int S(\lambda) T(\lambda) d\lambda + ZP,
\end{equation} 

where $mag$ is the ejecta magnitude provided by Suntzeff, and
$T(\lambda)$ is the filter transmission function.  $ZP$ is the
magnitude zero point given by \begin{math} ZP=2.5\log\int
S_{Vega}(\lambda) T(\lambda) d\lambda
\end{math} 
where $S_{Vega}(\lambda)$ is the flux of Vega derived from the Kurucz
stellar atmosphere models (Kurucz 1993) using $T_{eff}$ = 9,400~K and
log $g$ = 3.9.  Adopting 0.0 as the $JHK$ magnitude of Vega (Elias
 et al.  1982), $S(\lambda)$ was determined for each band and
epoch.  The scaling factor, $\phi$, was then obtained from
\begin{math} S(\lambda)=\phi \times S_{obs.}(\lambda)\end{math}. \\

\noindent To check the flux calibration procedure described above, we
carried out the following test.  The day~1734 $IJ$ spectra were
acquired under photometric conditions with good seeing, in a wide
(5.8~arcsec) slit. Therefore it was reasonable to assume that the
whole of the SN~1987A light and an uncertain amount of the Star~3
light were blended within the slit, and that the flux calibration via
the spectrophotometric standard was reliable.  We then edited out all
the circumstellar lines from the reduced, standard-fluxed spectrum,
yielding an unvignetted ejecta spectrum, but still blended an unknown
(vignetted) fraction of the Star~3 continuum.  We then scaled and
subtracted a blackbody representation of Star~3, with the scaling
adjusted so that, when multiplied by the CTIO filter function, the
resulting SN~1987A ejecta spectrum matched the CTIO photometry.  We
found that the flux of the ejecta spectrum produced in this way
differed from that obtained by the ``flat-continuum'' procedure by
less than 5\%.  We concluded that the ``flat-continuum'' procedure was
reliable, and so we applied the method to all the spectra.  The
factors by which the continuum-flattened spectra had to be scaled to
match the photometry ranged from $\times$0.3 to $\times$2.3,
indicating the severity of the effects discussed above.  Including
uncertainties in the Star~3 subtraction, we estimate the final
absolute fluxing of the ejecta spectra to be accurate to better than
$\pm$40\%.  The final spectra are shown in Figures~1--7. \\

\noindent We now turn briefly to the final fluxing of the CSM lines.
We first consider the CSM spectra obtained in the higher resolution
echelle mode.  While uncertainties in the continuum was not a problem
here, it was important to consider the effect of the extent of the
ring ($\sim$1.2~arcsec in the dispersion direction).  The echelle $IJ$
spectra were obtained with a wide slit (5.8~arcsec). In addition, all
spectra were extracted from a region of $>$3.2 arcsec. It is likely,
therefore, that any vignetting of the CSM flux was small.  Possible
narrow line contamination from the Be Star~3 was checked by inspection
of unpublished NICMOS observations of this star taken on day~4130
(P. Challis \& P. Garnavich, private communication). This showed that
the contribution to the narrow He~I 1.083~$\mu$m line was less than
12\%. We concluded that narrow line contamination was negligible.
Thus, the errors in the $IJ$ CSM line fluxes would be mostly of the
same origin as those present in the post-flattened ejecta spectra
({\it e.g.} variable extinction).  Consequently we corrected the $IJ$
CSM line fluxes using the same scaling factors, $\phi$, as for the
ejecta spectra.  We judge the precision in the final $IJ$ CSM spectra
to be better than $\pm$40\%.  Owing to S/N problems (see above), the
echelle $HK$ spectra were obtained with a narrow slit (1.4~arcsec).
Together with the effect of the typical 1.0--1.5~arcsec seeing, and
pointing errors, this meant that quite significant vignetting of the
CSM light probably occurred in the $HK$ observations.  We judge that
as much as half the CSM light could have been ``lost'' in this
way. Thus, while we also applied the $\phi$ scaling factor to the $HK$
CSM spectra, the effects of vignetting means that the $HK$ CSM fluxes
are very uncertain.  For the last two epochs (days~2952 and 3158), not
only was a narrow (1.6~arcsec) slit used, but the low resolution of
the grism produced strong blending of the CSM and ejecta
spectra. Consequently, no attempt has been made to extract CSM line
fluxes for these last two epochs.

\begin{table*}
\caption{Log of optical spectroscopy of SN~1987A}
\begin{minipage}{\linewidth}
\renewcommand{\thefootnote}{\thempfootnote}
\renewcommand{\tabcolsep}{0.13cm}
\begin {tabular}{llcccl}
\hline 
Date  &  Epoch & Spectral & Spectral &  Slit Width &    Seeing  \\
        &  (days) & Range (\AA) & Resn. ($\lambda/\Delta\lambda$) & (arcsec)  &
(arcsec) \\
\hline 
02 Nov 1990 & 1348 & 10690-13400  & 220  & 5.9 & $>$1.5 \\
03 Mar 1991 & 1469 & 8955-15120   & 385  & 5.8 & 1.0-3.0 \\
06 May 1991 & 1533 & 14320-24450  & 350  & 1.4 & $\sim$1.0 \\
23 Nov 1991 & 1734 & 14550-24670  & 350  & 1.4 &  $>$1.0\\ 
24 Nov 1991 & 1735 & 9035-15110   & 270  & 5.8 &  $\sim$1.0 \\  
19 Feb 1992 & 1822 & 14720-24500  & 350  & 1.4 & $\sim$1.5\\
20 Feb 1992 & 1823 & 9005-15050   & 270  & 5.8 &  1.0-3.0 \\
05 Dec 1992 & 2112 & 9048-14326   & 185  & 5.8 & $\sim$1.0\\
05 Dec 1992 & 2112 & 14690-24320  & 350  & 1.4 & $\sim$1.0\\
25 Mar 1995 & 2952 & 12250-21134  & 120  & 1.6 & 1.3-2.5 \\
17 Oct 1995 & 3158 & 12250-21110  & 120  & 1.6 & $\sim$1.2 \\
\hline
\end{tabular}
\end{minipage}
\label{log}
\end{table*}

\begin{figure*}
\vspace{16cm}
\includegraphics{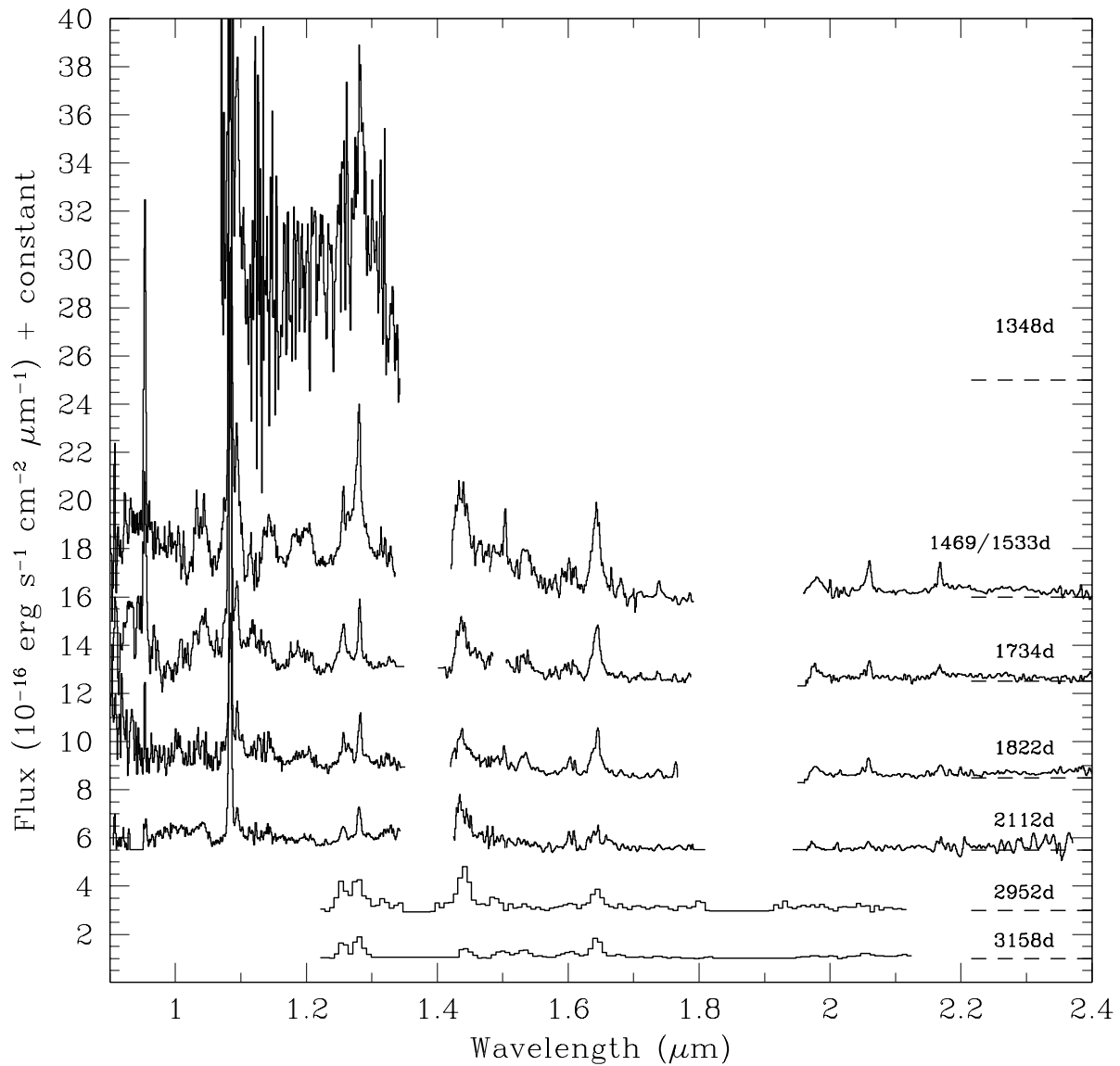}   
\caption{$IJHK$ band spectra of SN 1987A obtained with the IRIS
spectrograph at the Anglo-Australian Telescope.  For clarity the
spectra have been displaced vertically. Zero flux for each spectrum is
shown by the horizontal dashed lines on the right hand side.}
\label{figtot}
\end{figure*}

\begin{figure*}
\vspace{0.92\textheight}
\includegraphics{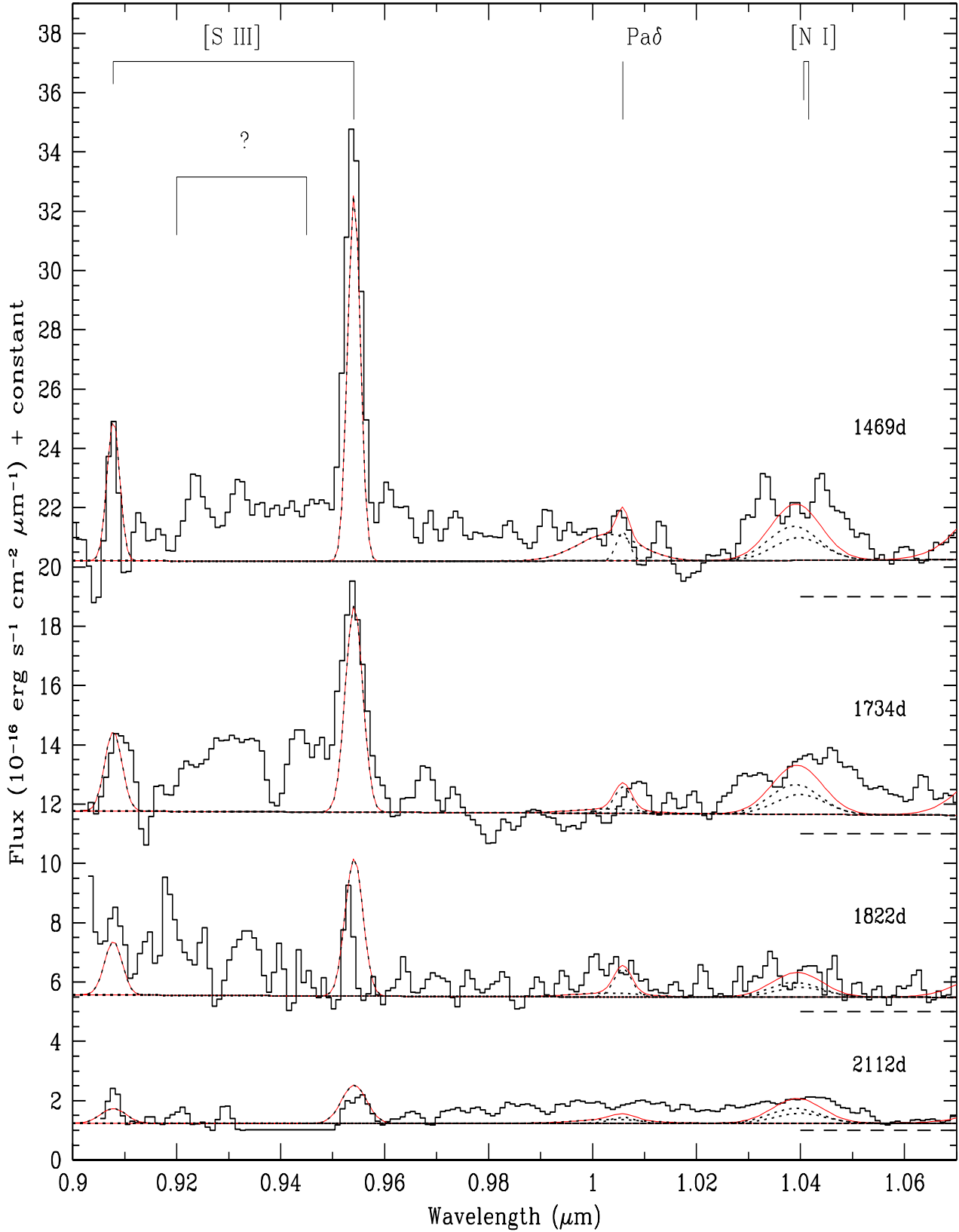}
\caption[]{$I$ band spectra (thick solid lines) of SN~1987A obtained
with the IRIS spectrograph in echelle mode, at the Anglo-Australian
Telescope.  For clarity the spectra have been displaced
vertically. Zero flux for each spectrum is shown by the horizontal
dashed lines on the right hand side.  The fluxes of the ejecta and
$IJ$-band CSM spectra are estimated to be accurate to better than
$\pm$40\%.  The thin solid lines show the empirical spectral
model (see text). The individual line components are shown as dotted
lines.  Line identifications are given by the vertical markers at the
top, placed at the rest wavelength for the SN~1987A
centre-of-mass. Within a multiplet, the lengths of the markers are
proportional to the component intensities (see text).}
\label{figi}
\end{figure*}

\begin{figure*}
\vspace{0.92\textheight}
\includegraphics{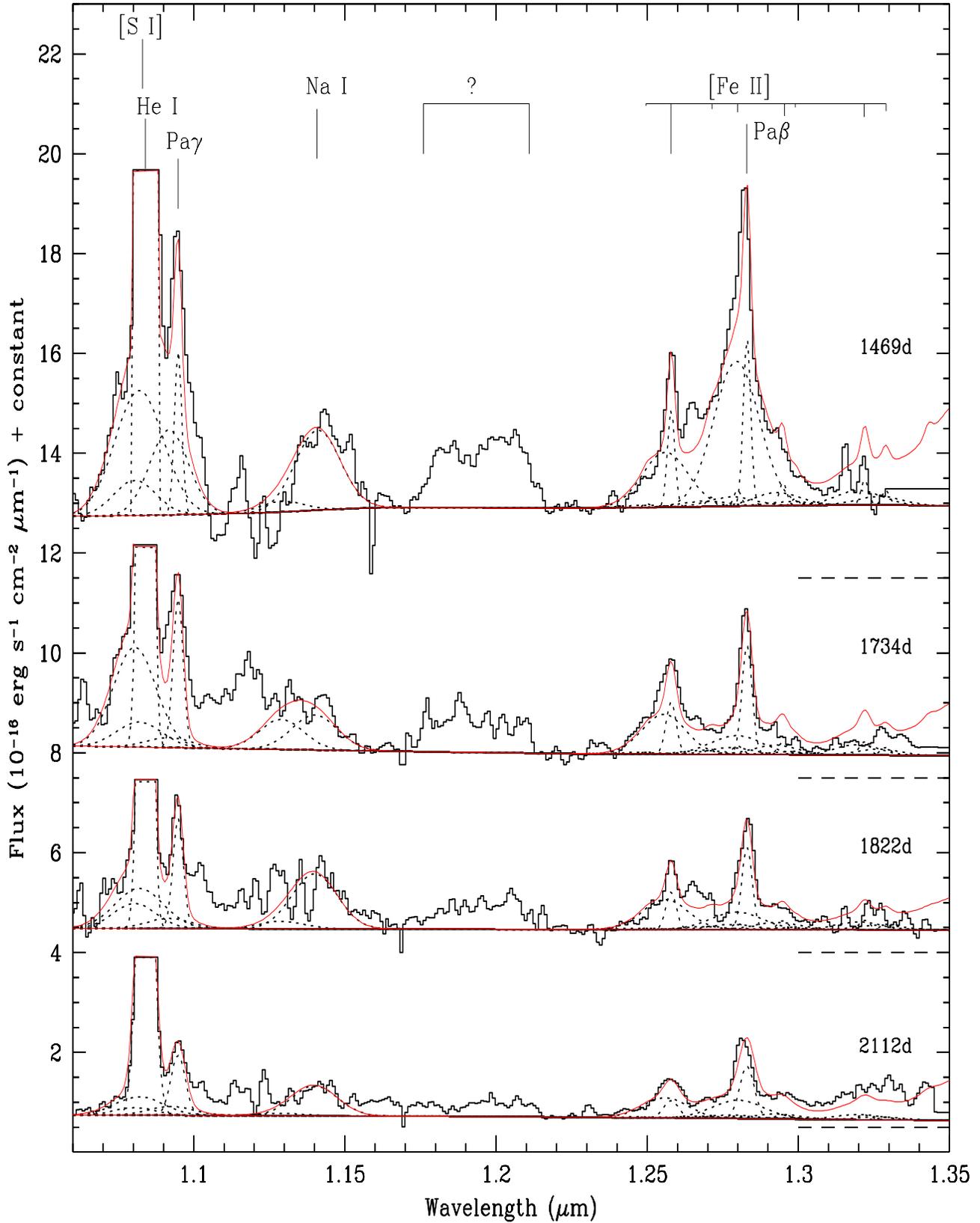} 
\caption{As Fig.~2, but for the $J$ window.  The strong He~I
1.083~$\mu$m CSM line has been truncated.}
\label{figj}
\end{figure*}

\begin{figure*}
\vspace{0.92\textheight}
\includegraphics{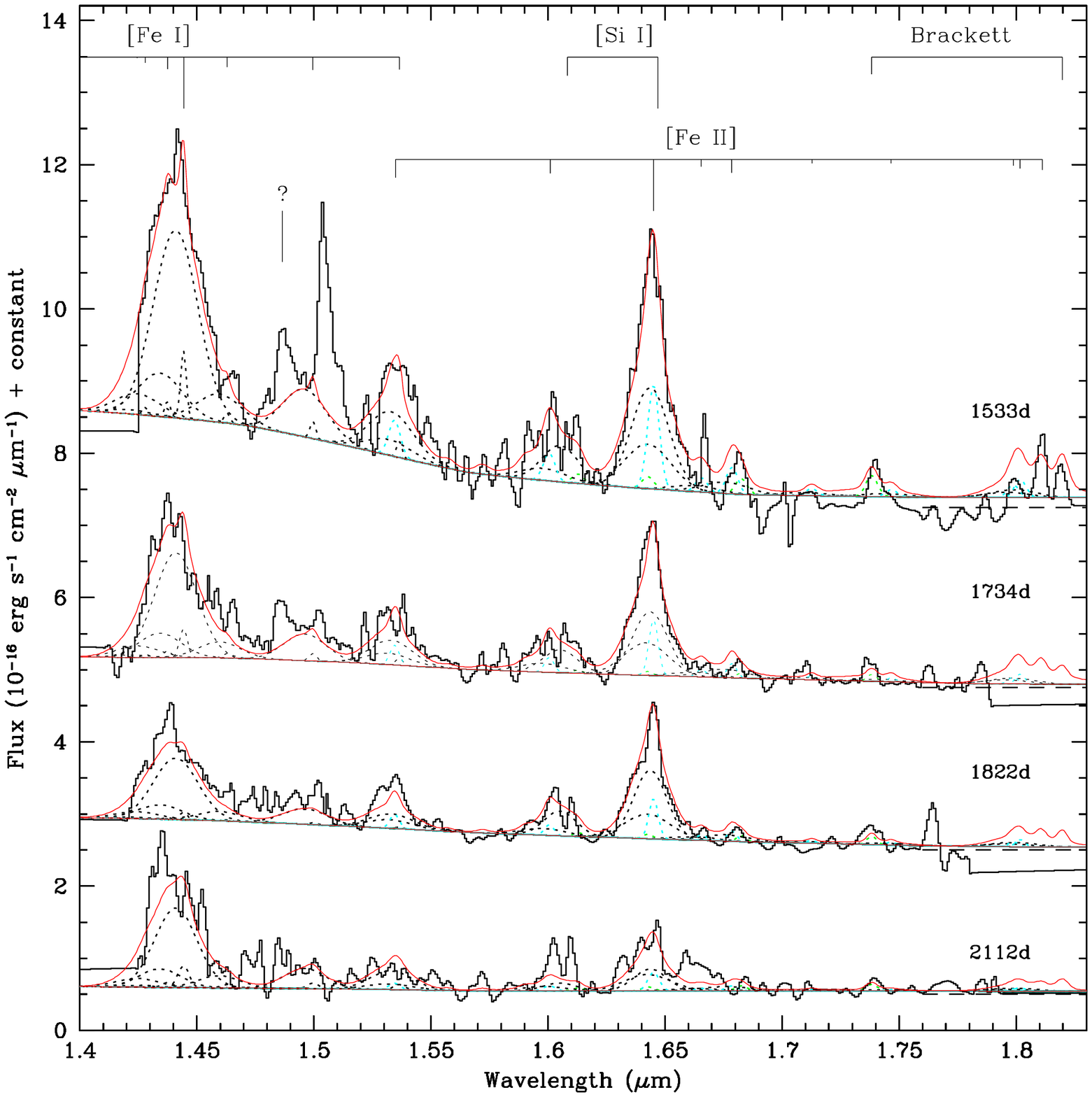}   
\caption{As Fig.~2, but for the $H$ window. Fluxing of the $HK$ CSM
lines is only approximate (see text).}
\label{figh}
\end{figure*}

\begin{figure*}
\vspace{0.92\textheight}
\includegraphics{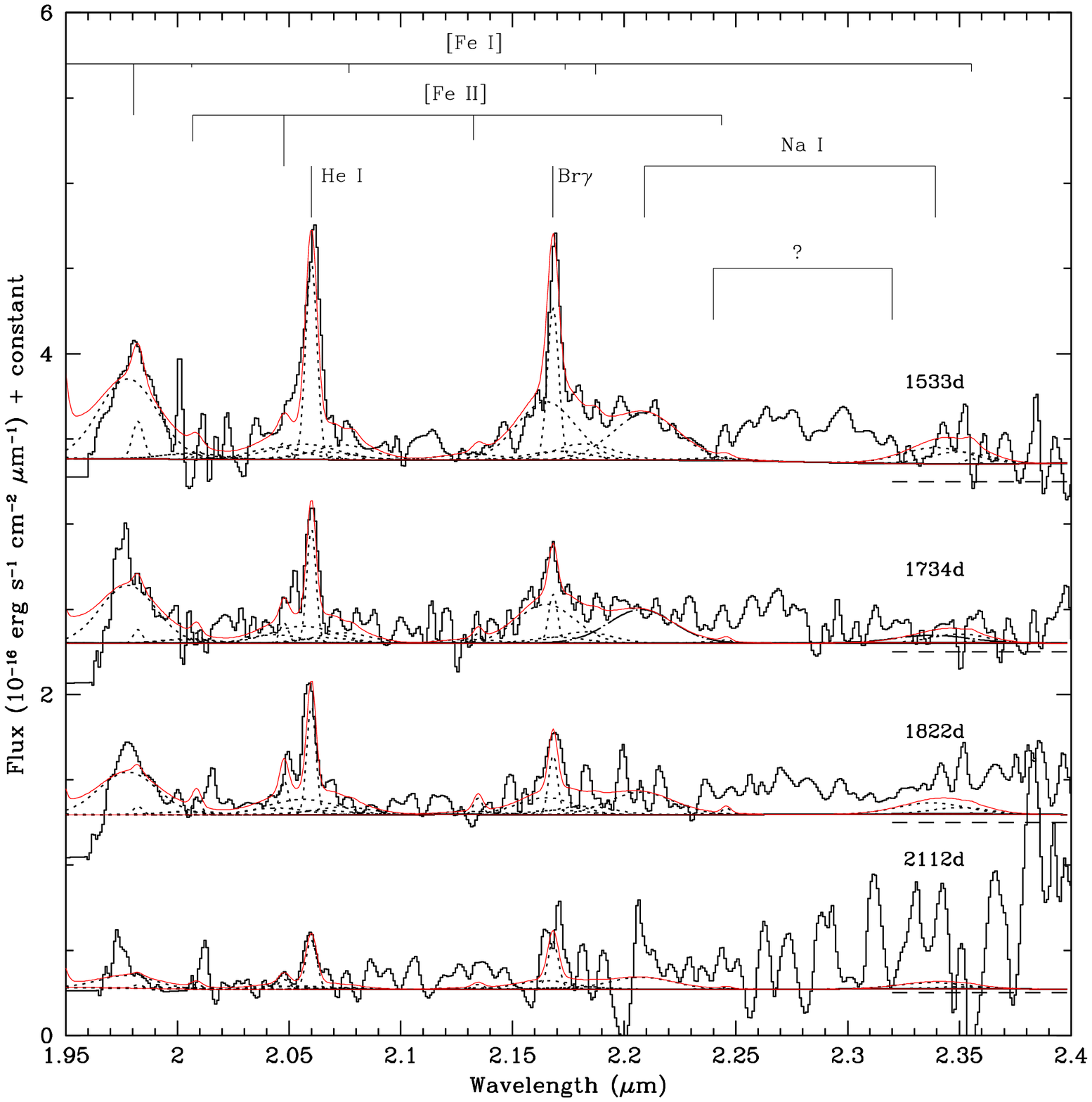}  
\caption{As Fig.~2, but for the $K$ window.  Fluxing of the $HK$ CSM
lines is only approximate (see text).}
\label{figk}
\end{figure*}

\begin{figure*}
\vspace{10.7cm}
\includegraphics{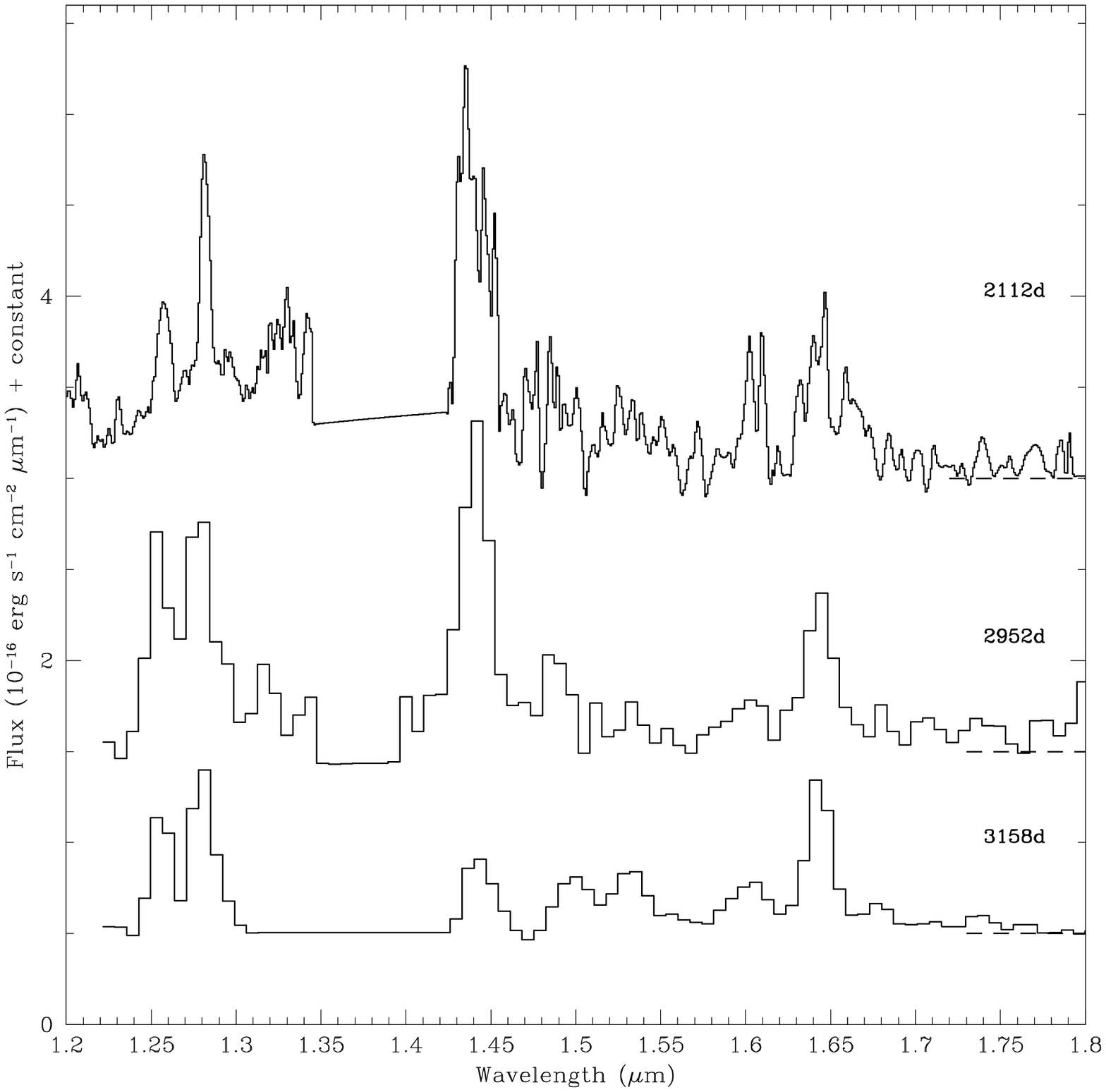}
\caption{IRIS grism-mode spectra of SN~1987A obtained on days~2952 and 3158.
Also shown is the last IRIS echelle spectrum (day~2112). For clarity the
spectra have been displaced vertically. Zero flux for each spectrum is
shown by the horizontal dashed lines on the right hand side.}
\label{fig_gr}
\end{figure*}

\section{Overview of the spectra}
The spectra (Figs.~1--7) comprise broad resolved emission lines from
the ejecta and narrow unresolved emission lines from the circumstellar
ring, superimposed on a continuum.  The continuum is roughly flat
within most wavelength bands, but fades over the range 1.4--1.7~$\mu$m
by a factor of $\sim$10 on day~1533, and $\sim$2--4 on days~1734, 1822
and 2122.  As already indicated, we attribute the continuum to a blend
of weak lines produced by the down-conversion of UV photons following
absorption/emission by metals.

\subsection{Line identification and measurement}
While it was straightforward to identify and measure the strongest or
most isolated lines (e.g. the He~I 1.083~$\mu$m CSM line), most of the
features comprised blends of lines from the ejecta and CSM and/or from
different species.  Therefore, in order to identify and measure the
component lines, we proceeded as follows.  Preliminary ejecta line
identification was performed by selecting plausible candidate species
from the line identifications of Paper~II and from the theoretical
predictions of dKLM98 and Kozma \& Fransson (1998a,b)
(KF98a,b). Likewise, we used the work of Lundqvist \& Sonneborn (2001)
to provide suggestions for the CSM line identifications.  Then, in
order to confirm the line identifications and to measure the
intensities, widths and wavelength shifts, we matched an empirical
spectral model to the data.  In this model, the continuum was
represented by applying a smoothed spline fit to regions of the
observed spectra which were clear of discrete emission
features. Emission features were then superimposed on this continuum.
These were selected using the aforementioned identification sources.
Their profiles were represented using Gaussian profiles with the peak
value as one of the free parameters. \\

\noindent For the CSM lines, we assumed that the lines were redshifted
by +289~km/s (Heathcote et al. 1991), and that their widths were defined
by the prevailing spectral resolution.  This resolution was measured
from the strong He~I 1.083~$\mu$m CSM line for the $IJ$ spectra, and
the He~I 2.058~$\mu$m CSM line for the $HK$ spectra ({\it cf.}
Section~2).  Within the [Fe~II] multiplets (see below), the relative
line intensities were derived by assuming that the population
distribution within the upper term was determined by the Boltzmann
distribution at 3000~K.  While the CSM temperature was probably
somewhat higher than this, it would have a negligible effect on the
distribution within the small energy range of a term.  For the [S~III]
CSM lines, direct measurement of the line intensities was judged to be
more reliable than use of the model matches.  \\

\noindent The empirical model ejecta line intensities, wavelength
positions and widths were all allowed to vary under certain
constraints.  For a particular species ({\it e.g.}  [Fe~II]),
individual linewidths and wavelength shifts were always set to be the
same in velocity space.  For the [Si~I] and [S~I] multiplets (see
below), the components of each multiplet had the same upper level and
so the intensity ratios of a given multiplet were found simply from
the A-values and statistical weights alone.  Given the uncertainty in
the excitation conditions, determination of some of the intensity
ratios within the [Fe~I] and [Fe~II] multiplets was less
straightforward since some lines arose from different upper levels.
In practice, to estimate the intensities of the other components we
simply used the same multiplet ratios as for the CSM, adopting
T=3000~K.  While this may have produced inappropriate relative
intensities for some lines, it is likely that most of the weaker lines
made only a minor contribution to the spectrum. We also note that for
the most significant line ratio {\it viz.}  $I_{1.257\mu
m}$/$I_{1.644\mu m}$, the components (which are actually members of
different multiplets) arose from the same upper level, and so could be
estimated from the A-values and statistical weights alone. Matching
was guided primarily by the absolute intensity of the dominant
component in a given multiplet.  For each observed spectrum, the
spectral model was varied until a visually satisfactory match to the
observed spectrum was achieved. \\

\noindent The model matches for epochs 1469 to 2112 are shown in
Figures~2--5.  Also shown (dotted lines) are the individual lines that
make up each blend.  It can be seen that some of the spectral features
comprise quite complicated blends of ejecta and CSM lines.  Details of
the line parameters for the ejecta and CSM are given in Tables~2 and 3
respectively.  In Table~2, Column~3 gives the displacement (in
velocity space) of the line peak with respect to the wavelength of the
identified line in the SN~1987A rest frame (redshift = +289~km/s).
Column~4 gives the linewidths (FWHM), again in terms of equivalent
velocity.  The precision in the absolute intensities of the
well-matched lines approaches that of the spectral fluxing precision
(Section~2), with even higher accuracy for relative intensities within
a single-band spectrum.  However, in many cases the matching is more
approximate with consequently poorer line intensity accuracy.  This
tends to be an increasing problem at later epochs.  To assess the
precision of a line intensity given in Table~1 or 2, the reader is
advised to examine the quality of the model matches in the spectral
plots (Figs.~2--5).  Model matches were carried out only up to
day~2112.  The spectra for days~2952 and 3158 were judged to be of too
low resolution and S/N to provide useful constraints on the model
parameters. We therefore simply present these spectra, in Figure~6,
compared with the final echelle spectrum taken on day~2112.

\begin{table*}
\caption{Major ejecta emission features observed on days 1469 to 2112}
\begin{minipage}{\linewidth}
\begin {tabular}{llllllr}
\\
Epoch & $\lambda _{peak}$& Line shift & Line width & Intensity& Identification & $\lambda
_{rest}$~ \\  
(days)    &   ($\mu$m) & (km/s)   &  FWHM  &   ($10^{-16}$erg   & &  ($\mu$m)~ \\
          &            &           & (km/s) &     s$^{-1}$cm$^{-2}$)      & \\ 
\\
1469   &  1.0805 & -400  & 4550 & 120 &  [S~I]~3p$^4$ $^3$P--3p$^4$ $^1$D(1F) & 1.0820  \\    
1734   &  1.0805 & -400  & 4550 & 350 &                &        \\
1822   &  1.0805 & -400  & 4550 &  90 &                &        \\
2112   &  1.0805 & -400  & 4550 &  25 &                &        \\
\\   			
1469   &  1.0820 & -550  & 5200 & 500 &   He~I~2s$^3$S--2p$^3$P$^{\rm o}$ & 1.0830 \\  
1734   &  1.0820 & -550  & 5200 &  95 &                &   \\
1822   &  1.0820 & -550  & 5200 & 160 &                &         \\
2112   &  1.0820 & -550  & 5200 &  75 &                &         \\
\\    		
1469   &  1.0918 & -800  & 4150 & 270 &   H I Pa$\gamma$ & 1.0938  \\     
1734   &  1.0918 & -800  & 4700 &  35 &               &        \\  
1822   &  1.0918 & -800  & 4700 &  35 &                &        \\
2112   &  1.0922 & -700  & 4700 &  35 &                &        \\
\\ 
1469   &  1.1290 & -400  & 4550 &  40 & [S~I]~3p$^4$ $^3$P--3p$^4$ $^1$D(1F) & 1.1306  \\    
1734   &  1.1290 & -400  & 4550 & 110 &                &        \\
1822   &  1.1290 & -400  & 4550 &  30 &                &        \\
2112   &  1.1290 & -400  & 4550 &   8 &                &        \\
\\   			                            
1469   &  1.1415 &   60  & 4650 & 300 & Na~I~3p$^2$P$^{\rm o}$--4s$^2$S~(3) & 1.1397  \\
1734   &  1.1415 & -200  & 4650 & 130 &                &        \\
1822   &  1.1415 & -200  & 4650 & 210 &                &        \\
2112   &  1.1415 & -200  & 4650 & 115 &                &        \\
\\    				                            
1469   &  1.2560 & -450 & 4200 & 210 & [Fe~II]~a$^6$D$_{9/2}$--a$^4$D$_{7/2}$  & 1.2567 \\
1734   &  1.2555 & -570 & 4050 & 145 &                &        \\
1822   &  1.2560 & -450 & 4200 & 115 &                &        \\
2112   &  1.2560 & -450 & 4200 &  75 &                &        \\
\\
1469   &  1.2790 & -800 & 4400 & 550 & H I Pa$\beta$  & 1.2818  \\	
1734   &  1.2800 & -800 & 4700 &  80 &                &        \\
1822   &  1.2790 & -800 & 4700 &  75 &                &        \\
2112   &  1.2800 & -700 & 4700 &  80 &                &        \\
\\   		   				                        
1469   & 1.4410 & -700 & 4650 & 620  &[Fe~I]~a$^5$D$_4$--a$^5$F$_5$ & 1.4430 \\
1734   & 1.4410 & -700 & 4400 & 330  &       &  \\
1822   & 1.4410 & -700 & 4650 & 200  &                &        \\
2112   & 1.4410 & -700 & 4400 & 250  &                &        \\       
\\    				                            
1469   & 1.6051 & -600 & 3400 &  100 & [Si~I]~3p$^3$P$_1$--3p$^1$D$_2$~(0.01F) & 1.6068 \\ 
1734   & 1.6051 & -600 & 3200 &  60 &                &        \\ 
1822   & 1.6051 & -600 & 3000 &  60 &                &        \\
2112   & 1.6051 & -600 & 3000 &  20 &                &        \\
\\
1469   & 1.6437 & -600 & 3400 & 280 & [Si~I]~3p$^3$P$_2$--3p$^1$D$_2$ (0.01F) & 1.6454  \\
1734   & 1.6437 & -600 & 3200 & 165 &                &       \\
1822   & 1.6437 & -600 & 3000 & 165 &                &        \\
2112   & 1.6437 & -600 & 3000 &  55 &                &        \\
\\    				                            
1469   & 1.6426 & -450 & 4200 & 150 & [Fe~II]~a$^4$F$_{9/2}$--a$^4$D$_{7/2}$ & 1.6435  \\
1734   & 1.6420 & -550 & 4050 & 100 &                &        \\
1822   & 1.6426 & -450 & 4200 &  85 &                &        \\
2112   & 1.6426 & -450 & 4200 &  55 &                &        \\ 
\\  
\end{tabular}
\label{ideje}
\end{minipage}
\begin{minipage}{\linewidth}
NOTE: Uncertainties in Line shifts and Line widths are typically
$\pm$200~km/s and $\pm$500~km/s respectively. The precision in the
absolute intensities of the well-matched lines is about $\pm$35\%,
with higher accuracy for relative intensities within a single-band
spectrum.  However, where the matching is poorer the error in the line
intensities is consequently larger. This tends to be an increasing
problem at later epochs.  To assess the precision of a line intensity
given in the Table, the reader is advised to examine the quality of
the model matches in the spectral plots (Figs.~2--5).
\end{minipage}
\end{table*}
\begin{table*}
\contcaption{}
\begin{minipage}{\linewidth}
\begin {tabular}{lllllll}
\\ 
Epoch & $\lambda _{peak}$&  Line shift & Line width & Intensity& Identification & $\lambda_{rest}$~ \\  
(days)    &  ($\mu$m)  &  (km/s)   &  FWHM  &   ($10^{-16}$erg   & &  ($\mu$m)~ \\
          &            &           &  (km/s)  &     s$^{-1}$cm$^{-2}$)     & \\  
\\
1469   & 1.9777  & -700 & 4650 & 155 & [Fe~I]~a$^5$F$_5$--a$^3$F$_4$ &  1.9804  \\
1734   & 1.9777  & -700 & 4400 & 110 &                &        \\
1822   & 1.9777  & -700 & 4650 &  80 &                &        \\
2112   & 1.9777  & -700 & 4400 &  25 &                &        \\
\\    		                            
1469   & 2.0446  & -450 & 4200 &  20 & [Fe~II]~a$^4$P$_{5/2}$--a$^2$P$_{3/2}$ & 2.0457  \\
1734   & 2.0437  & -550 & 4050 &  15 &                &        \\
1822   & 2.0446  & -450 & 4200 &  10 &                &        \\
2112   & 2.0446  & -450 & 4200 &   6 &                &        \\     
\\
1469   & 2.0561  & -550 & 5200 &  35 & He~I~2s$^1$S--2p$^1$P$^{\rm o}$ & 2.0580  \\
1734   & 2.0561  & -550 & 4900 &  35 &                &        \\
1822   & 2.0561  & -550 & 5200 &  35 &                &        \\
2112   & 2.0561  & -550 & 5200 &   8 &                &        \\
\\		                            
1469   & 2.1650  & -450 & 4250 & 110 & H I Br$\gamma$ & 2.1661  \\
1734   & 2.1650  & -450 & 4250 &  75 &                &        \\
1822   & 2.1650  & -450 & 4250 &  35 &                &        \\
2112   & 2.1650  & -450 & 4250 &  15 &                &        \\ 
\\
1469   & 2.2095  &   60 & 4650 & 100 & Na~I~4s$^2$S--4p$^2$P$^o$& 2.2070  \\
1734   & 2.2075  & -200 & 4650 &  70 &                &        \\
1822   & 2.2075  & -200 & 4650 &  50 &                &        \\
2112   & 2.2075  & -200 & 4650 &  25 &                &        \\ 
\\
\end{tabular}					     
\label{ideje}
\end{minipage}
\begin{minipage}{\linewidth}
NOTE: Uncertainties in Line shifts and Line widths are typically
$\pm$200~km/s and $\pm$500~km/s respectively. The precision in the
absolute intensities of the well-matched lines is about $\pm$35\%,
with higher accuracy for relative intensities within a single-band
spectrum.  However, where the matching is poorer the error in the line
intensities is consequently larger. This tends to be an increasing
problem at later epochs.  To assess the precision of a line intensity
given in the Table, the reader is advised to examine the quality of
the model matches in the spectral plots (Figs.~2--5).
\end{minipage}					     
\end{table*}					     
						     
\begin{table}
\caption{Major CSM emission features observed on days 1469 to 2112}
\begin{minipage}{\linewidth}
\renewcommand{\thefootnote}{\thempfootnote}
\begin {tabular}{lllr}
\\ 
Epoch & Intensity& Identification & $\lambda
_{rest}$~ \\  
(days)    &   ($10^{-16}$erg   & &  ($\mu$m)~ \\
          &    s$^{-1}$cm$^{-2}$)    & & \\
\\
1469   & 120(s) &  [S~III]       &  0.9069 \\
1734   & 135(s) &                &         \\
\\    		      		                            
1469   & 490(s) &  [S~III]       &  0.9532 \\
1734   & 265(s) &                &         \\
\\    				                            
1348   &3900 & He~I~2s$^3$S--2p$^3$P$^{\rm o}$ & 1.0830   \\
1469   &3600 &                &         \\
1734   &1200 &                &         \\
1822   & 900 &                &         \\
2112   & 530 &                &          \\
\\    				                            
1348   & 280  & H I Pa$\gamma$ & 1.0938  \\
1469   & 100 &                &         \\
1734   & 124 &                &        \\
1822   &  95 &                &        \\
2112   &  60 &                &        \\
\\    				                            
1469   &  60 & [Fe~II]~a$^6$D$_{9/2}$--a$^4$D$_{7/2}$  & 1.2567  \\
1734   &  45 &                &        \\
1822   &  30 &                &        \\
2112   &  20 &                &        \\
\\		                            
1348   & 340 & H I Pa$\beta$  & 1.2818  \\
1469   & 100 &                &        \\
1734   &  90 &                &        \\
1822   &  70 &                &        \\
2112   &  65 &                &        \\
\\
1469   &  90 & [Fe~II]~a$^4$F$_{9/2}$--a$^4$D$_{7/2}$ & 1.6435  \\
1734   &  40 &                &        \\
1822   &  30 &                &        \\
2112   &  15 &                &        \\ 
\\    				                            
1469   &   6 & [Fe~II]~a$^4$P$_{5/2}$--a$^2$P$_{3/2}$ & 2.0459  \\
1734   &   6 &                &        \\
1822   &  10 &                &        \\
2112   &   4 &                &        \\
\\
1469   &  70 & He~I~2s$^1$S--2p$^1$P$^{\rm o}$ & 2.0581  \\
1734   &  35 &                &        \\
1822   &  30 &                &        \\
2112   &  20 &                &        \\
\\		                            
1469   &  55 & H I Br$\gamma$ & 2.1661  \\
1734   &  15 &                &        \\
1822   &  20 &                &        \\
2112   &  15 &                &        \\
\\
\end{tabular}					     
\label{idcsm}
\end{minipage}
\begin{minipage}{\linewidth}
NOTE: The precision in the absolute intensities of the well-matched
$IJ$-band lines is about $\pm$35\%, with higher accuracy for relative
intensities within a single-band spectrum.  However, where the
matching is poorer the error in the line intensities is consequently
larger. This tends to be an increasing problem at later epochs.  To
assess the precision of a line intensity given in the Table, the
reader is advised to examine the quality of the model matches in the
spectral plots (Figs.~2--5).  Vignetting problems in the $HK$-band CSM
spectra (see text) means that the error in intensities is considerably
larger than in the $IJ$-band (see text).  (s) denotes that the fluxes
were measured directly from the observations.
\end{minipage}					     
\end{table}

\subsection{The ejecta lines}
For the epochs covered by the echelle spectra (days~1469--2112), the
most prominent ejecta emission features included H~I (Pa$\beta$,
Pa$\gamma$, Br$\gamma$), He~I 1.08, 2.06~$\mu$m, Na~I 1.14, 2.21,
2.34~$\mu$m, [Si~I] 1.61, 1.65~$\mu$m, [Fe~I] 1.44, 1.98~$\mu$m and
[Fe~II] 1.26, 1.64, 2.05~$\mu$m.  [S~I] 1.08, 1.13~$\mu$m may also be
present, but strong blending with other lines renders this
identification less positive.  There were also regions of emission for
which we were unable to find plausible identifications.  These were at
0.92--0.95, 1.03--1.05, 1.17--1.21 and 2.24--2.32~$\mu$m.  The
1.03--1.05~$\mu$m feature may be caused partly by [N~I] 1.04~$\mu$m.
However, the width and shape of the feature indicates that there must
be contributions from other species.  The 1.17--1.21~$\mu$m feature is
particularly interesting in that it has persisted from as early as
day~112 (Paper~I). In Papers~I and II we attributed this feature to a
blend of allowed lines of Mg~I, Si~I and K~I. However, its relatively
unchanging shape between days~112 and 1822 prompts us to question if
it really is a blend of different species.  We also note that the
2.24--2.32~$\mu$m emission lies in the same wavelength region as an
unidentified feature discussed in Paper~II.

\subsection{The CSM lines}			     
The principal CSM features were due to H~I (Pa$\beta$, Pa$\gamma$,
Br$\gamma$), He~I 1.08, 2.06~$\mu$m, [S~III] 0.91, 0.95~$\mu$m and
[Fe~II] 1.26, 1.646, 2.05~$\mu$m.  Of particular note is the very
strong He~I 1.083~$\mu$m line which contributes $\sim$20\% of the
total (ejecta+CSM) $IJ$~band flux on day 1469 and 13-11\% on days
1734-2112 (Figure~7).  We believe that [S~III] 0.907, 0.953~$\mu$m
lines have never before been reported in a supernova spectrum.  Their
presence in the spectra of days~1469 and 1734 is particularly
convincing.  The lines are less apparent in the day~1822 and 2112
spectra due to the fading flux and the poorer observing conditions at
these epochs.  We have also included narrow line profiles for the
[Fe~I] 1.443, 1.98~$\mu$m (and the other lines of these
multiplets). This line was not anticipated in the work of Lundqvist \&
Sonneborn (2001).  However, the observed line profile shape on
day~1533 in these wavelength regions led us to suspect that CSM [Fe~I]
emission may be making a small contribution to these features. At
later epochs, the presence of [Fe~I] emission from the CSM is much
less convincing. Finally, we note the presence of an unidentified
narrow feature at a wavelength of 1.485~$\mu$m (after blueshifting by
289~km/s) on days~1533 and 1734.

\begin{figure*}
\vspace{10.7cm}
\includegraphics{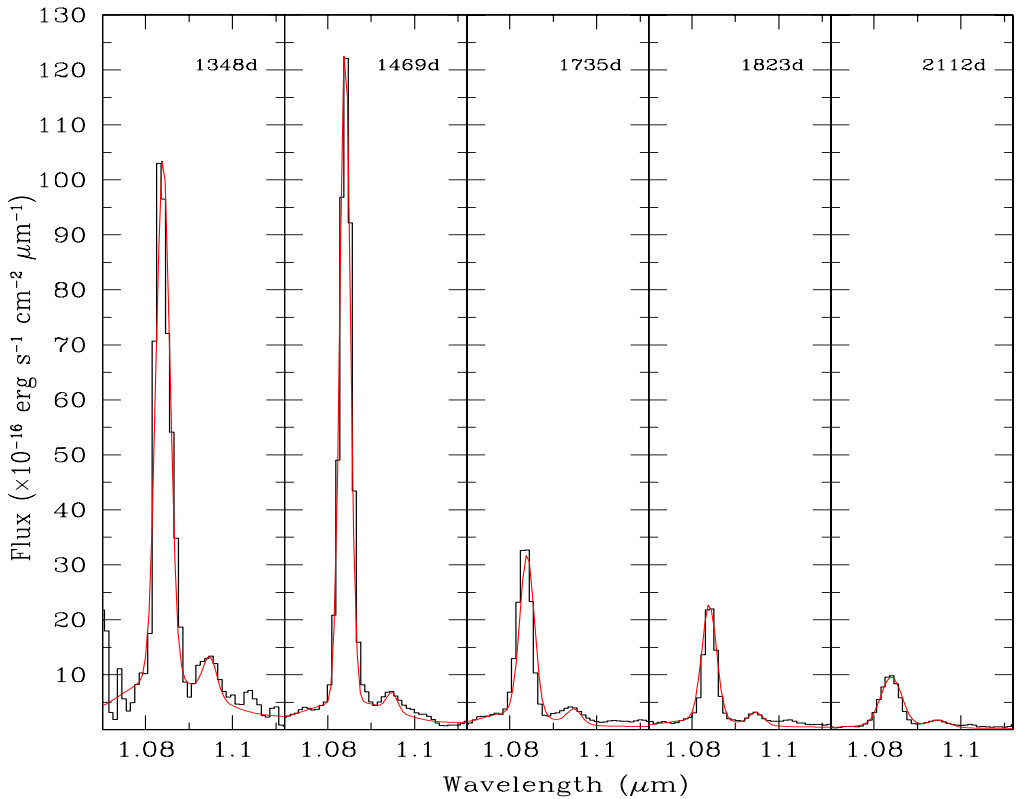} 
\caption{The evolution of the He~I 1.083~$\mu$m and Pa$\gamma$ CSM
lines in the spectra of SN~1987A.  The lines are always
unresolved. Differences in the FWHM are due to variation in the seeing
conditions.  This is the reason for the increase in peak flux between
days~1348 and 1469.  However, the integrated intensities of the lines
show a monotonic decline (see Table~3). The fluxing of the
CSM spectra is better than $\pm$40\%. }
\label{fig_hecsm}
\end{figure*}

\section{Discussion}						     
\subsection{The Physical Framework}
The CSM spectra will be discussed elsewhere.  Here we confine our
discussion to the ejecta spectra. The framework for this section is
based upon theoretical descriptions of the physical conditions in the
ejecta and line strength predictions at very late times provided by
Kozma \& Fransson (1992), Fransson \& Kozma (1993), Fransson, Houck \&
Kozma (1996) (FHK96), C97, dKLM98, and KF98a,b.  While dKLM98 evolve
their model to only 1200~days ({\it i.e.} 150~days prior to the
beginning of our observations) their results are still relevant to the
earlier part of the era considered here.  KF98a follow the temperature
and ionisation evolution of their model to 2000~days, covering a
significant portion of our era, but their line emission calculations
(KF98b) stop at 1200~days.  C97 consider the much later epoch of
day~2875, which is near the end of our era. \\

\noindent During the era of the observations described here, the
SN~1987A ejecta spectrum is produced in a rather exotic manner.  The
energy source of the nebula is the radioactive decay of a number of
species created in the explosion (Woosley, Pinto \& Hartmann 1989;
Timmes et al. (1996)).  These include $^{56}$Co, $^{57}$Co, $^{60}$Co,
$^{44}$Ti and $^{22}$Na.  All emit $\gamma$-rays.  In addition,
$^{56}$Co, $^{44}$Ti and $^{22}$Na emit energetic positrons and
$^{60}$Co emits energetic electrons.  By day~1348, the most
significant energy sources were $^{57}$Co and $^{44}$Ti, contributing
about one half and one quarter respectively of the deposited energy
(Li, McCray \& Sunyaev 1993; Timmes et al. 1996).  Around day~1565,
the dominant source of deposited energy became $^{44}$Ti and by about
day~2000 more than half the deposited $^{44}$Ti energy was in the form
of positrons.  It is generally assumed that positrons are deposited
``on the spot''.  Consequently, given the $\sim60$~year lifetime of
$^{44}$Ti, this means that once positron deposition is dominant, the
rate of energy deposition approaches a constant value of
$1.3\times10^{36}$ erg/s per $10^{-4}$~M$_\odot$ of synthesised
$^{44}$Ti (Woosley et al. 1989). \\

\noindent In order to reproduce the observed spectra and evolution of
SN~1987A at late times, C97, dKLM98 and KF98a,b begin with model
nebulae having a number of zones of differing chemical composition.
These zones are arranged so as to reflect the strong, deep macroscopic
mixing in SN~1987A for which there is a great deal of evidence ({\it
e.g.} Arnett 1988, Pinto \& Woosley 1988, Shigeyama, Nomoto \&
Hashimoto 1988, Woosley 1988, Haas  et al. 1990, Spyromilio,
Meikle \& Allen 1990, Fassia \& Meikle 1999).  Light elements (H, He)
are mixed down to low velocities, and iron group elements are mixed up
to high velocities.  C97, dKLM98 and KF98a,b introduce this mixing in
different ways.  However, they all invoke a macroscopically mixed
``core'' lying within $\sim$2000~km/s. (dKLM98 refer to the inwardly
mixed H/He zones as the ``inner envelope''.)  The core thus contains
zones that are H-rich, He-rich, intermediate-element-rich, and
iron-group-rich, with the nebula being bathed in the radioactive decay
energy.  The core is generally of mass 4--6~M$_\odot$. In addition,
C97 and KF98a,b include an outer H-envelope of mass 10~M$_\odot$
extending out to a velocity of 6000--7000~km/s. \\

\noindent In general, the fraction of the radioactive decay energy
that does not directly escape from the nebula is injected into the
nebular material via Coulomb interaction with the positrons and
Compton-scattered electrons.  The energy of the resulting non-thermal
high energy electrons then goes towards excitation, ionisation and
heating of the nebula (KF98a).  Three effects play important roles in
the evolution of the temperature and ionisation at the very late times
considered here.  These are the ionisation/thermal ``freeze-out''
effect, adiabatic cooling, and the ``IR-catastrophe''.  The first two
are important in the H/He envelope, while the third plays an important
role in the metal-rich core. 

\subsubsection{Freeze-out}
The ionisation freeze-out effect was suggested originally by Clayton
 et al. (1992) and by Fransson \& Kozma (1993).  These authors
pointed out that as the SN evolves, there will eventually come a time
when the recombination timescale exceeds the radioactive or expansion
timescale, so that the rate of change in the level of ionisation slows
significantly.  Once this phase is reached, the bolometric luminosity
exceeds that of the instantaneous radioactive decay deposition, since
some of the luminosity results from recombination following ionisation
at a significantly earlier epoch.  In the model of KF98a,b the
ionisation freeze-out phase begins at 800--900~days in most zones, but
has a more pronounced effect on the luminosity in the H/He envelope.
However, in their ``inner-envelope'' model, dKLM98 find that the
freeze-out does not occur until much later than this.  Instead, they
identify a ``thermal freeze-out'' where the radiative cooling timescale
exceeds that of the expansion timescale.  In their model, this occurs
before the ionisation freeze-out but, as with ionisation freeze-out,
results in a luminosity exceeding that of the instantaneous
radioactive decay deposition.  There is general agreement that by
day~1200, the ejecta are predominantly neutral, but with a fraction of
singly-ionised species resulting from the ionisation freeze-out and/or
direct ionisation by non-thermal electrons.  During 1350--2000~days
the electron fraction is $\sim10^{-3}$ in the H-envelope, rising to
0.1 in the Fe-He zone (KF98a). At 3425~days, Lundqvist  et al.
(2001) calculate that the fraction of iron that is singly-ionised, is
in the range 0.2--0.4. 

\subsubsection{Adiabatic cooling} 
The adiabatic cooling becomes significant when the radiative cooling
timescale becomes longer than the expansion timescale.  For pure
adiabatic cooling, T$\propto t^{-2}$.  Owing to their lower density
and metallicity, adiabatic cooling is already dominant as early as
$\sim$250~days in the H-envelope and $\sim$800~days in the He-envelope
(KF98a).  For the H/He zones within the core, the higher densities
mean that adiabatic cooling begins to dominate at 800-1000~days.
While fine-structure line cooling ({\it cf.} IR catastrophe below)
becomes the dominant radiative cooling mechanism in these zones, it
never supersedes adiabatic cooling.  Nevertheless, fine-structure line
cooling may be significant in these regions (C97).  KF98a find that
the H-envelope temperature lies in the range 400-1000~K on day~1350,
falling to a range of 150--300~K by day 2000. The temperatures of the
H/He zones within the core fall from $\sim$900~K to $\sim$300~K over
the same period.  The day~2000 KF98a value compares well with the
$\sim$300~K derived from the Balmer continuum by Wang  et al.
(1996) for about the same time.  The model of C97 yields $\sim$130~K
for the H-envelope at 2875~days, compared with 350$\pm$100~K derived
from the Balmer continuum at that time. 

\subsubsection{The IR catastrophe}
Once the nebular heating/cooling rate drops below a certain level,
cooling via low-lying fine-structure transitions dramatically
overtakes optical and near-IR transitions as the dominant radiative
cooling mechanism.  Consequently the stabilising temperature falls
abruptly from $\sim$2000~K to a few $\times100$~K, and the bulk of the
nebula's luminosity shifts to far-IR emission.  This effect is known
as the ``IR Catastrophe'', and was predicted by Axelrod (1980) in his
pioneering work on type~Ia spectral models.  Fransson \& Chevalier
(1989) predicted that it could also occur in core-collapse SNe such as
SN~1987A.  The first direct evidence for this phenomenon occurring in
SN~1987A was obtained through the detailed study of the evolution of
the near-IR/optical [Fe~II] lines during the second year (Spyromilio
\& Graham 1992).  \\

\noindent In the dense, metal-rich zones, fine-structure radiative
cooling dominates for a considerable time.  [Fe~II]~26~$\mu$m emission
is particularly important.  Fine-structure cooling is most pronounced
in the Fe-rich zones (KF98a), where it commences at $\sim$500~days at
2700~K.  According to KF98a, by the beginning of our observations
(1348~days) the temperature in the Fe-He zone had fallen to
$\sim150$~K, where it remained until 2000~days, after which adiabatic
cooling became important. The IR catastrophe is also important for the
intermediate mass element zones, where the temperature at 1350~days is
200--400~K, falling to 100--200~K by day~2000 (KF98).  Both FK98b and
dKLM98 find that, owing to the IR-catastrophe, [Fe~II] 0.72, 1.26,
1.53, 1.64~$\mu$m lines originating from the newly-synthesised iron in
the core should have vanished by $\sim$600~days.  However, detectable
flux in these lines persists beyond this epoch.  FK98b and dKLM98
attribute this to thermally-excited emission from primordial iron in
the H-He envelope where the temperature remains above $\sim$2000~K
until after day~1000. However, as indicated above, by 1350~days, the
temperature in the H-He zone is below 1000~K and so thermally excited
emission from these lines should be undetectable. \\
					    
\subsection{Near-IR ejecta lines at very late times}
We can use the observed near-IR spectra to examine the physical
scenario described above.  The evolution of the intensities of the
principal ejecta lines is illustrated in Figure~8.

\begin{figure*}
\vspace{13cm}
\includegraphics{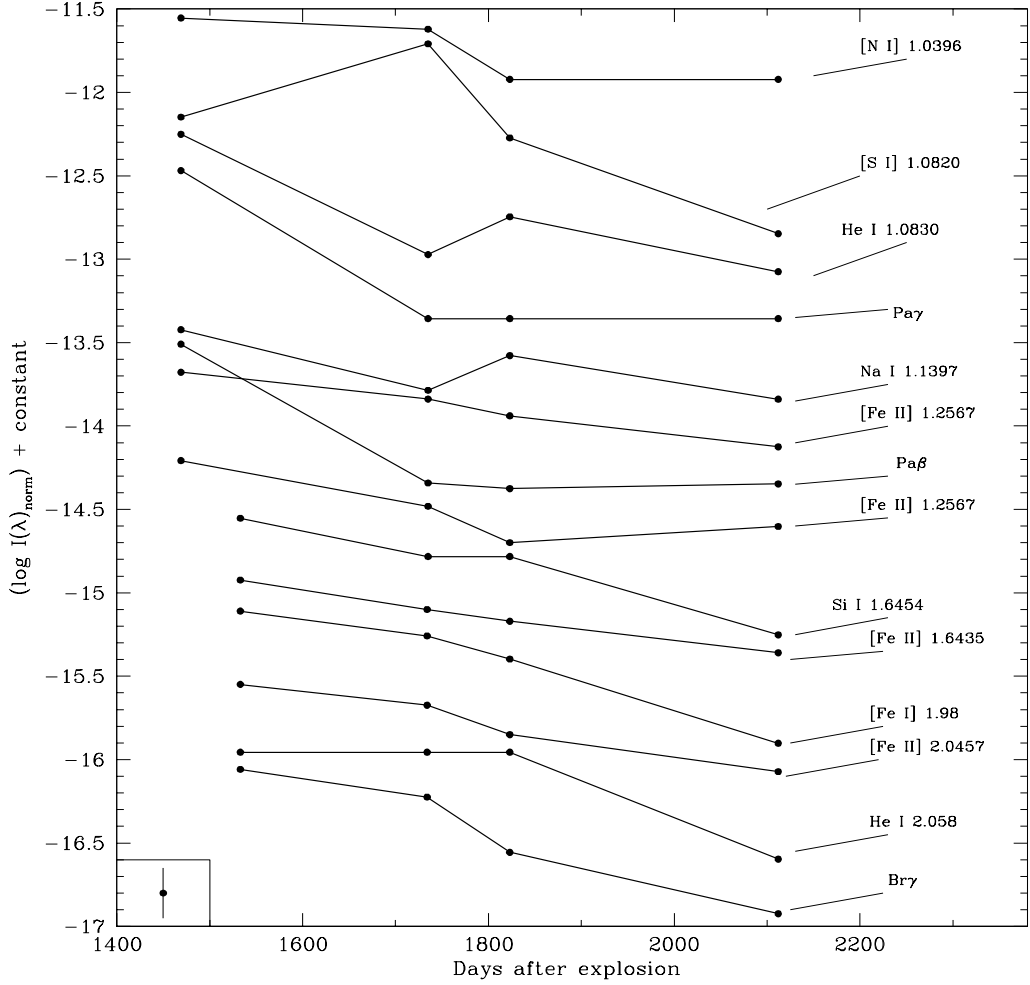} 
\caption{Evolution of the line intensities of the most prominent
ejecta lines. The boxed point with the error bars indicates the
typical $\pm$35\% fluxing precision of the ejecta spectra.  However,
where the spectral model matching is poor, the error in intensity is
consequently greater (see text).}
\label{fig_evol}
\end{figure*}

\subsubsection{Hydrogen Lines: Pa$\beta$, Pa$\gamma$, Br$\gamma$} 
The intensity and evolution of the H-lines in the adiabatic cooling
phase is calculated by dKLM98 and KF98b up to 1200~days.  For epochs
$>2$~years, non-thermal ionisation from the ground state by
Compton-scattered electrons dominates photoionisation from excited
states (C97, KF98b). Non-thermal excitations are also important at
epochs beyond 1000~days (KF98b).  dKLM98 find that by day~800 the line
ratios should be Case~B. FK98b reach a similar conclusion, but with
the transition to Case~B in the core being a little later, at about
1000~days, being complete by day~1200.  Thus, during our era, the
H-spectrum is predicted to consist of a mixture of a low-temperature
Case~B H-recombination spectrum together with a contribution from
direct non-thermal excitation (see KF98b for details).  Line light
curves are given by dKLM98 and KF98b for the near-IR lines
Pa$\alpha$,$\beta$, Br$\alpha$,$\beta$,$\gamma$,$15$.  dKLM98 find
that, for the well-observed near-IR H-lines Pa$\beta$ and Br$\gamma$,
their models makes satisfactory predictions to 1200~days. This is in
spite of their non-inclusion of the envelope with its delayed
recombination effects.  The KF98b model light curve for Br$\gamma$,
which includes delayed recombination effects in the envelope, is also
in good agreement with the observations.  Their model indicates that,
by day~1200, the H-envelope is responsible for $\sim$70\% of the
Br$\gamma$ flux, and that this fraction is growing.  Both Fransson et
al. (1996) and dKLM98 point out that derivation of total H-mass from
the H-lines is difficult and subject to large uncertainties. Fransson
et al. (1996) and KF98b make use of the line profiles in order to find
the most appropriate model (and hence the mass of H). (In a sense, C97
also make use of the observed velocity distribution in determining the
H-mass).\\

\noindent We examined the ejecta H-line fluxes to see if they yield
the predicted low-temperature Case~B ratios. These were first
de-reddened assuming $A_{V}=0.6$ (Blanco et al. 1987).  We find that
the observed, de-reddened intensity ratio
$I_{Pa\gamma}/I_{Pa\beta}=0.45-0.6$ for days~1469, 1734, 1822.  Martin
(1988) predicts a value of $\sim$0.5 for a temperature of 200--1000~K.
Given the uncertainty in the relative line intensities, we judge that
the observed ratios are consistent with the predicted values.  The
uncertainties in the model matching are too large to make a useful
judgement about the ratio on epoch day~2112.  Martin (1988) also
predicts $I_{Br\gamma}/I_{Pa\beta}$=0.195.  This is in fair agreement
with the observed, de-reddened ratio of 0.18 on day~1469.  Apparently
poorer agreement is obtained on subsequent days.  However, this is
probably again due to model matching uncertainties, especially in the
$K$-window.  We conclude that, at least up to $\sim$1800~days, the
observed H-spectra are consistent with the scenario of non-thermal
ionisation/excitation within an environment dominated by adiabatic
cooling.

\subsubsection{Helium Lines:  He~I 1.083, 2.058~$\mu$m}
As for hydrogen, theoretical considerations indicate that
post-1000 day He-lines evolve in an environment where the temperature
is dominated by adiabatic cooling, and the lines are driven by (a)
non-thermal direct excitation, and (b) recombination following
non-thermal ionisation.  dKLM98 and KF98b point out that the
2.058~$\mu$m line is uncontaminated by other ejecta species throughout
the observations, and also find that it is optically thin in all
regions after 700~days.  Moreover, after $\sim$700~days, the
2.058~$\mu$m line intensity is relatively insensitive to assumptions
made about the He~I 584~\AA\ continuum destruction probability
(KF98b). These considerations lead dKLM98 \& KF98b to suggest that the
post-600~day He~I 2.058~$\mu$m line can provide the best measure of
the total helium mass.  However, consideration of our spectra leads us
to a more pessimistic view of the usefulness of the 2.058~$\mu$m line
in this situation.  By day~1533, it was quite faint. To make matters
worse it is blended with comparably-strong CSM flux in this line.
Moreover, He~I 2.058~$\mu$m lies in a bad part of the atmospheric
window.  Consequently there is large uncertainty in the He~I
2.058~$\mu$m line flux.  dKLM98 show that their model 2.058~$\mu$m
light curve provides a good match to the observations up to 1100~days.
The KF98b light curve match is somewhat poorer, with a flux
overproduction of about 40\% between days 800 \& 1100.  The
2.058~$\mu$m light curves of both dKLM98 and KF98b stop at 1200~days,
but suggest a gradual slowing down.  For the later epochs described
here, to within the uncertainties, the observed fluxes fall on a
plausible extrapolation of the dKLM98 model light curve, but continues
to fall below that of the (extrapolated) KF98b light curve. \\

\noindent dKLM98 \& KF98b agree that the He~I 1.083~$\mu$m line is
more difficult to use for the determination of helium abundance.  It
is more temperature sensitive, and is optically thick for much longer
than is the case for He~I 2.058~$\mu$m.  Indeed, KF98b find the
1.083~$\mu$m line to be optically thick in the He~I region even beyond
2000~days. However, high optical depth in the inner zones may not be
too important since, after 1200~days, the 1.083~$\mu$m emission from
the H-zone is predicted to dominate (and was already optically thin at
700 days).  KF98b also point out that there may be contamination due
to [S~I]~1.082~$\mu$m. Our empirical model matches also indicate that
[S~I]~1.082~$\mu$m emission may be present. However, this line is part
of a multiplet, with another component lying at 1.13~$\mu$m and having
about 0.3 of the intensity. Examination of this part of the spectrum
(in spite of it being in a bad part of the atmospheric window)
indicates that [S~I]~1.082~$\mu$m makes, at most, a relatively minor
contribution to the 1.08~$\mu$m ejecta emission (see Fig.~3).  We find
that the observed fluxes of the He~I 1.083~$\mu$m line fall on
plausible extrapolations of both the dKLM98 and KF98b model light
curves.  

\subsubsection{Forbidden lines of neutral species: [Si~I], [S~I]
[Fe~I]} The low temperature (T$<$400~K) of the electron gas in the
macroscopically-mixed core during the period of our observations
implies that thermally-excited near-IR ejecta lines of [Si~I], [S~I]
\& [Fe~I] should have faded below detectability.  Yet, the [Si~I] and
[Fe~I] lines are quite clearly present during this very late phase.
(Strong blending makes the presence of the [S~I] lines more
ambiguous.) In particular, the [Si~I] 1.645~$\mu$m and [Fe~I]
1.443~$\mu$m lines are visible to beyond day~2000.  The persistence of
these lines provides valuable support for the proposition that, by
this era, these lines are produced either by recombination or through
direct excitation by non-thermal electrons. \\

\noindent The [Si~I]~1.6~$\mu$m multiplet is of particular interest
here.  FHK96 point out that for this species, recombination to the
neutral state does not produce any significant line emission in the
optical or near-IR region.  Consequently, the [Si~I]~1.6~$\mu$m
emission must be due entirely to direct excitation by non-thermal
electrons.  An interesting consequence of this is that the line
luminosity will follow the instantaneous energy input, and will be
independent of the ejecta temperature.  At very late times this is
dominated by the 100\% absorption of $^{44}$Ti decay positrons. Thus,
if the $^{44}$Ti scenario is correct, the luminosity in these lines
should converge to a near-constant value.  In Figure~8 we see that the
[Si~I] 1.645~$\mu$m line fades by about a factor of 3 between
days~1734/1822 and 2112.  During this period, the radioactive energy
deposition is dominated by $^{44}$Ti and would fade by a factor of
about 1.5.  Thus, to within the uncertainty in the line intensities,
the evolution of [Si~I] 1.645~$\mu$m is consistent with the
radioactive decay and energy deposition being dominated by $^{44}$Ti
decay at these late times.  We note that FHK96 also show that lines
which are driven purely by non-thermal excitation, such as the
[Si~I]~1.6~$\mu$m multiplet, have the potential to provide reliable,
temperature-insensitive mass estimates, provided the line profile and
the bolometric luminosity is also known. \\

\noindent C97 suggest that, by day~2875, virtually the entire positron
luminosity of the $^{44}$Ti is deposited in the Fe/Si-rich clumps.
While cooling is mostly by ground-term fine-structure lines, about
10\% (10$^{35}$~erg/s) is via the UV-optical-NIR lines of neutral
species. About 20\% (of the $10^{35}$ erg/s) is emitted as identified
UV lines of Fe~I and Si~I with a further $\sim$70\% being
down-converted to numerous allowed and forbidden optical/NIR metal
lines, forming a quasi-continuum.  Of the remaining $\sim10^{34}$
erg/s, a substantial fraction flows into the [Si~I] 1.6~$\mu$m and
[Fe~I] 1.44~$\mu$m near-IR multiplets.  This appears to be supported
by our low-resolution spectra (Figure~6) taken on days~2952 and 3158
where the [Fe~I] 1.443~$\mu$m and possibly the [Si~I] 1.645~$\mu$m
features are still detected. The luminosity in just the [Fe~I]
1.443~$\mu$m line on day~2952 is about $0.7\times10^{34}$ erg/s. 

\subsubsection{Forbidden Lines of singly-ionised iron}
As with the neutral forbidden lines, the low temperature of the ejecta
during the era considered here means that there would have been
negligible thermal excitation of the [Fe~II]~1.26, 1.64~$\mu$m
multiplets.  This includes the excitation of primordial iron in
the H-envelope since, as mentioned above, even there it is expected
that the temperature would be less than 1000~K by day~1350 (KF98a).
This suggests, therefore, that the persistence of the [Fe~II] lines
must be due to recombination or direct excitation by non-thermal
electrons.  However, we note that C97 state that near-IR [Fe~II] lines
produced by radiative cascade are expected to be weak since the major
radiative cascade to the ground level goes through optical forbidden
lines and FIR lines of the ground term.  Detailed modelling is
required to test if this is in conflict with our detection of these
lines.

\subsection{Velocity behaviour in the ejecta lines} 
The FWHM velocities of the more prominent ejecta lines are listed in
Table~2, Col.~4. The He~I lines apparently exhibit the largest width,
at $\sim$5000~km/s (FWHM).  However, the uncertainty in this
measurement is large due to strong blending of the He~I ejecta lines
with CSM lines and/or other species.  The H~I lines and the [Fe~I],
[Fe~II] lines show widths of 4000--4500~km/s, while the [Si~I] lines
have widths closer to 3000~km/s.  Even this latter value is higher
than the $\sim$2000~km/s invoked for the macroscopically mixed core.
Of particular interest is the fact that the [Fe~I] and [Fe~II] lines
exhibit higher velocity widths than the [Si~I] lines.  We argued above
that the persistence of the forbidden iron line emission during the
era studied here must be driven by delayed recombination following
freeze-out or through direct excitation by non-thermal electrons.
Consequently, the high velocities in these lines imply that these
processes must be occurring well out into the H/He envelope.  Detailed
modelling will be required to determine how much each process
contributes to the line emission.  If we favour the latter scenario,
it immediately suggests that upward mixing was even greater than has
been assumed hitherto.  We note that velocities of at least
$\sim$3000~km/s were observed in the [Fe~II] lines as early as the end
of year~1 (Paper~I and Spyromilio et al. 1990, Haas {\it et al.}
1990).  Moreover, Fassia \& Meikle (1999) showed that the presence of
the He~I 1.083~$\mu$m line on days~76--135 implied that upward mixing
of $^{56}$Ni in the ejecta of SN~1987A had extended to velocities as
high as 3000--4000~km/s.  Upward mixing of $^{56}$Ni to even higher
velocities (over 5000~km/s) has been recently deduced by Mitchell et
al. (2001) on the basis of the high strength of the Balmer lines a few
days after explosion.  Such high $^{56}$Ni velocities may be evidence
of neutrino-instability-driven acceleration of radioactive nickel just
after the core-bounce ({\it e.g.} Herant, Benz \& Colgate 1992).
Alternatively, these results may favour the jet-like explosion models
of Nagataki (2000), in which the high velocity [Fe~II] line profiles
are well reproduced. \\

\noindent Another remarkable characteristic of these late-time spectra
is the presence and persistence of blueshifts (with respect to the
supernova rest frame) in the ejecta lines (Table~2, Col.~3), with
values of typically --200~km/s to --800~km/s.  Such shifts first
appeared around day~600 (Paper~II) and were attributed to the
formation of dust, blocking out the red (far-side) wing of the line.
With the persistence of the blueshifts to as late as day~2000 we
conclude that very dense dust concentrations must have formed in the
ejecta. However it is possible that asymmetry in the excitation
conditions may also be contributing to the effect.

\section{Summary}
We have presented near-IR spectra of SN~1987A covering the period 3.7
and 8.6~years post-explosion.  This is the first time that IR spectra
of a supernova has been obtained to such late epochs. We have
described the measures taken to remove contamination from the nearby
Stars 2 and 3.  The resulting spectra comprise emission from both the
ejecta and the bright circumstellar ring.  The contributions from
these two sources were separated out using an empirical spectral
model.  The CSM emission lines comprise recombination lines of H~I and
He~I, and forbidden lines of [S~III] and [Fe~II].  The allowed ejecta
spectra include lines of H~I, He~I, Na~I, [Si~I], [Fe~I], [Fe~II] and
possibly [S~I].  We are unable to confirm the presence of weak [Co~I]
and [Co~II] lines as suggested by Bautista  et al.  (1995) on
day~1445. \\

\noindent The intensity ratios and widths of the H~I ejecta lines are
consistent with the predicted low-temperature Case~B recombination
spectrum arising from non-thermal ionisation/excitation in an
extended, adiabatically cooled H-envelope.  Owing to difficulties with
low signal-to-noise and CSM-ejecta blending, the He~I 2.058~$\mu$m
ejecta line is probably of less value than pure ejecta theory would
suggest.  \\

\noindent Perhaps the most interesting result is the slow decline in
the ejecta forbidden lines.  This is particularly important for the
[Si~I] lines, since it supports the scenario that pure non-thermal
excitation was taking place, driven increasingly predominantly by the
decay of $^{44}$Ti. The data presented here offers the prospect of
improved abundance measurements of silicon in the ejecta. \\

\noindent The FWHM of the ejecta lines provide evidence that the
extensive mixing has occurred, with heavy elements reaching layers
with velocities as high as $\sim$3000-4500~km/s. The iron velocities
were particularly high, and may provide support for the
neutrino-instability fast-nickel scenario, or for a jet-like
explosion.  The blueshifts of the ejecta lines that had appeared
around day 600 (Paper II) continued to be present. This probably
indicates that dense concentrations of dust persisted in the ejecta
even as late as day~2000. However, asymmetry in the excitation
conditions may also contribute to the observed blueshifts.\\

\noindent The spectra presented here are unique.  It is likely to be
many years before such late-time near-IR observations are obtained for
another supernova.  Moreover, with the increasing interaction of the
ejecta with the circumstellar ring, these near-IR spectra are probably
the last from the ``pristine ejecta'' of SN~1987A that can be
obtained.

\section{Acknowledgements}
We dedicate this paper to the memory of the late David Allen.  He
played a central role in the near-IR study of SN~1987A.  He made
particularly vital contributions to the work described here, both as
project scientist for IRIS, and by carrying out most of the
observations.  He is sorely missed. \\

\noindent We thank Mike Burton, Stuart Lumsden, Raylee Stathakis and
Chris Tinney for carrying out some of the observations, and the staff
of the AAO for their support.  We also thank Nick Suntzeff, Peter
Challis and Peter Garnavich for the use of their unpublished data, and
Emma Bowers, Robert Cumming and Gian Varani for assistance with the
data reduction.  This work is supported through PPARC Grant
No. PPA/G/S/1997/00266.

\end{document}